\documentclass[11pt]{article}
\usepackage{moriond,epsfig}

\def\Journal#1#2#3#4{{#1} {\bf #2}, #3 (#4)}

\def\PLB{{\em Phys. Lett.}  B}
\def\PRL{\em Phys. Rev. Lett.}
\def\PRD{{\em Phys. Rev.} D}

\def\be{\begin{equation}}
\def\ee{\end{equation}}
\def\bea{\begin{eqnarray}}
\def\eea{\end{eqnarray}}

\begin{document}
\vspace*{4cm}
\title{\boldmath{$\Upsilon$}(5S) AND \boldmath{$B_s$} DECAYS AT BELLE}

\author{A. DRUTSKOY}

\address{Department of Physics, University of Cincinnati, 345 Clifton ct., \\
 Cincinnati, OH 45221, USA}

\maketitle\abstracts{
Recent results obtained using the data sample of 23.6\,fb$^{-1}$
collected on the $\Upsilon$(5S) resonance with the Belle detector
at the KEKB asymmetric energy $e^+ e^-$ collider are discussed.
Measurements of several $B_s^0$ decay branching fractions are reported.
Studies of the $\Upsilon$(5S) decays to the channels with $B^+$ and $B^0$ 
mesons or bottomonium states are discussed.}

\section{Introduction}

During the last several years an opportunity for $B_s^0$ meson studies at the 
$e^+ e^-$ colliders running at the $\Upsilon$(5S) resonance 
has been extensively explored. The first evidence for $B_s^0$ production at the
$\Upsilon$(5S) was found by the CLEO collaboration \cite{cleoi,cleoe} using
a dataset of 0.42\,fb$^{-1}$ collected in 2003. 
To test the feasibility of a $B_s^0$ physics program the Belle collaboration 
collected at the $\Upsilon$(5S) a dataset of 1.86\,fb$^{-1}$ in 2005. 
After the successful analysis of these data \cite{beli,bele}, Belle collected 
a bigger sample of 21.7\,fb$^{-1}$ in 2006. More $\Upsilon$(5S) data were
taken by Belle in 2008.

The $\Upsilon$(5S) resonance can potentially decay to various final states
with $B^+$, $B^0$ and, moreover, $B_s^0$ mesons, because
the $\Upsilon$(5S) has a mass exceeding the
$B_s^0\bar{B}_s^0$ production threshold.
At this energy a $b\bar{b}$ quark pair can be produced and 
hadronized in various final states,
which can be classified as
two-body $B_s^0$ channels $B_s^0\bar{B}_s^0$, 
$B_s^0\bar{B}_s^\ast$, $B_s^\ast\bar{B}_s^0$, $B_s^\ast\bar{B}_s^\ast$,
two-body $B^{+/0}$ channels
$B\bar{B}$, $B\bar{B}^\ast$, $B^\ast\bar{B}$, $B^\ast\bar{B}^\ast$,
three-body channels $B\bar{B}\,\pi$, $B\bar{B}^\ast\,\pi$, 
$B^\ast\bar{B}\,\pi$, $B^\ast\bar{B}^\ast\,\pi$, and
four-body channel $B\bar{B}\,\pi \pi$.
Here $B$ denotes a $B^+$ or a $B^0$ meson and 
$\bar{B}$ denotes a $B^-$ or a $\bar{B}^0$ meson, and
the excited states decay to their ground states via
$B^\ast \to B\gamma$ and $B_s^\ast \to B_s^0\gamma$.
Moreover a $b\bar{b}$ quark pair can hadronize to a bottomonium
state accompanied by $\pi$, $K$, or $\eta$ mesons, for example through
the $\Upsilon$(5S)$\to \Upsilon$(1S)$\,\pi^+\pi^-$ decay.
Fractions and decay parameters for all of these channels
provide important information about the $b$-quark dynamics.

The $B_s^0$ production rate at the $\Upsilon$(5S) was measured to be 
$(19.5^{+3.0}_{-2.3})\%$ \cite{pdg}, therefore the $\Upsilon$(5S) could 
play a similar role for comprehensive $B_s^0$ studies that the 
$\Upsilon$(4S) has played for $B^+$ and $B^0$ studies. 
Similar to the experimental technique used at the $\Upsilon$(4S), 
two variables can be used to identify $B_s^0$ signals at the $\Upsilon$(5S):
the energy difference $\Delta E\,=\,E^{CM}_{B_s^0}-E^{\rm CM}_{\rm beam}$
and the beam-energy-constrained mass
$M_{\rm bc} = \sqrt{(E^{\rm CM}_{\rm beam})^2\,-\,(p^{\rm CM}_{B_s^0})^2}$,
where $E^{\rm CM}_{B_s^0}$ and $p^{\rm CM}_{B_s^0}$ are the energy and momentum
of the $B_s^0$ candidate in the $e^+ e^-$ center-of-mass (CM) system,
and $E^{\rm CM}_{\rm beam}$ is the CM beam energy.
The $B_s^* \bar{B}_s^*$, $B_s^* \bar{B}_s^0$,
$B_s^0 \bar{B}_s^*$ and $B_s^0 \bar{B}_s^0$ intermediate channels 
can be distinguished kinematically in the $M_{\rm bc}$ and $\Delta E$ plane,
where three well-separated $B_s^0$ signal regions
can be defined corresponding to the cases where both, only one, or neither
of the $B_s^0$ mesons originate from a $B_s^*$ decay.

\section{\boldmath{$B_s^0$} decay branching fraction measurements}

\subsection{Measurement of $B_s^0 \to D_s^- \pi^+$ decay and evidence for $B_s^0 \to D_s^{\mp} K^{\pm}$ decay}\label{subsec:dspi}

We report here the results from studies of
$B_s^0 \to D_s^- \pi^+$ and $B_s^0 \to D_s^{\mp} K^{\pm}$ decays \cite{bdspi}
obtained by the Belle collaboration with 23.6\,fb$^{-1}$ at the $\Upsilon$(5S).
In this analysis
$D_s^-$ candidates are reconstructed in the $\phi \pi^-$, $K^{*0} K^-$
and $K^0_S K^-$ modes.
The $M_{\rm bc}$ and $\Delta E$ scatter plots for the
$B_s^0 \to D_s^- \pi^+$ and $B_s^0 \to D_s^{\mp} K^{\pm}$ decays are studied.
A clear signal is observed in the $B_s^0 \to D_s^- \pi^+$
decay mode, and evidence for the $B_s^0 \to D_s^{\mp} K^{\pm}$ decay
is also seen.
For each mode, a two-dimensional unbinned extended maximum likelihood
fit in $M_{\rm bc}$ and $\Delta E$ is performed on the selected candidates.
Fig.\,1 shows the $M_{\rm bc}$ and $\Delta E$ projections
in the $B_s^* \bar{B}_s^*$ region of the data, together with the fitted
functions.
The different fitted components are shown with dashed curves for the
signal, dotted curves for the $B_s^0 \to D_s^{*-} \pi^+$ background,
and dash-dotted curves for the continuum.

\begin{figure}
\begin{center}
\psfig{figure=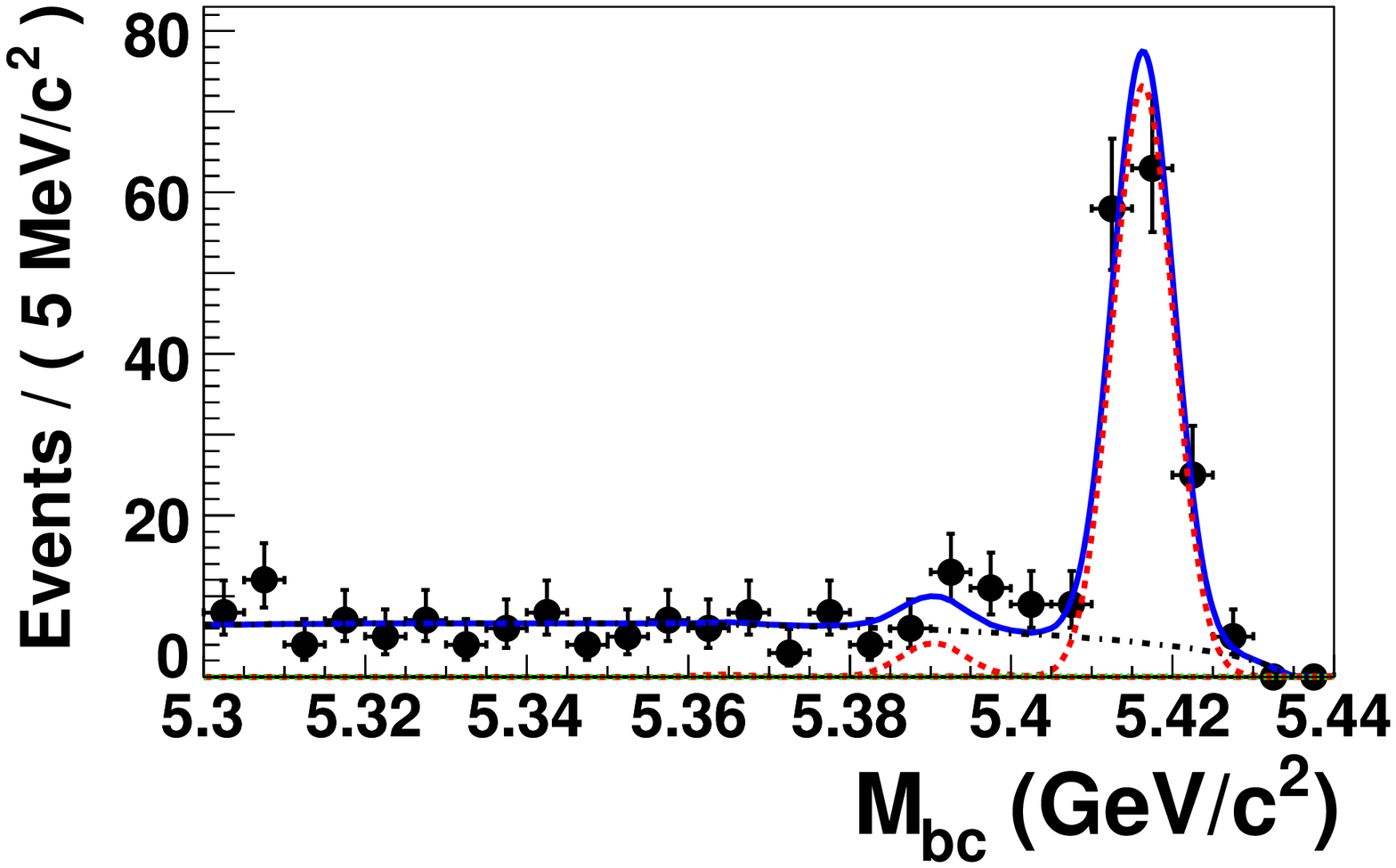,height=1.5in}\psfig{figure=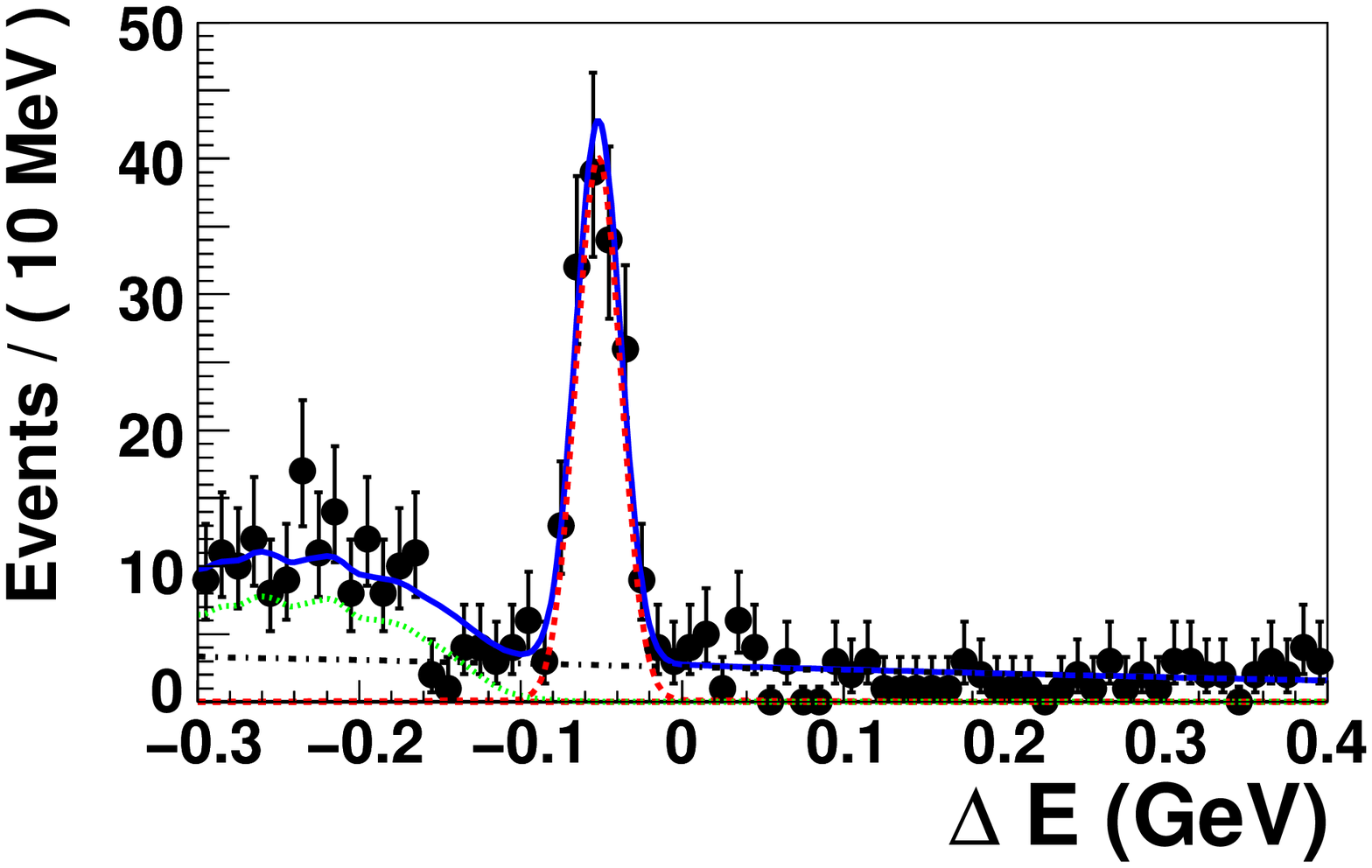,height=1.5in}
\end{center}
\vspace{-0.4cm}
\caption{The $M_{\rm bc}$ distribution (left) and $\Delta E$ distribution (right) for the $B_s^0 \to D_s^- \pi^+$ candidates.
\label{fig:dspib}}
\end{figure}

Finally, the branching fractions 
${\cal B}(B_s^0 \to D_s^- \pi^+) = (3.67^{+0.35}_{-0.33} \pm 0.65) \times 10^{-3}$ and 
${\cal B}(B_s^0 \to D_s^{\mp} K^{\pm}) = (2.4^{+1.2}_{-1.0} \pm 0.4) \times 10^{-4}$ are measured.
The ratio ${\cal B}(B_s^0 \to D_s^{\mp} K^{\pm}) / {\cal B}(B_s^0 \to D_s^- \pi^+) = (6.5^{+3.5}_{-2.9})\%$ is derived; the errors are completely
dominated by the low $B_s^0 \to D_s^{\mp} K^{\pm}$ statistics.

Comparing the number of events reconstructed in the 
$B_s^0 \to D_s^- \pi^+$ mode in three signal regions, 
the fraction of $B_s^\ast\bar{B}_s^\ast$
events over all $B_s^{(\ast)}\bar{B}_s^{(\ast)}$ events
was measured to be
$f_{B_s^\ast\bar{B}_s^\ast}=(90.1^{+3.8}_{-4.0} \pm 0.2)\%$.
From the $B_s^0$ signal fit the mass
$m(B_s^*) = (5416.4 \pm 0.4 \pm 0.5)\,$MeV/$c^2$ is obtained.
The mass difference $m(B_s^*)-m(B_s^0)$ 
obtained is 4.0$\sigma$ larger than the world average for
$m(B^{*0})-m(B^0)$.

\subsection{Observation of $B_s^0 \to \phi \gamma$ and search for $B_s^0 \to \gamma \gamma$ decays}\label{subsec:gaga}

The $B_s^0 \to \phi \gamma$ and $B_s^0 \to \gamma \gamma$ decays \cite{bfiga}
are studied by Belle with 23.6\,fb$^{-1}$ at the $\Upsilon$(5S).
Within the Standard Model the $B_s^0 \to \phi \gamma$ decay 
can be described
by a radiative penguin diagram and the
corresponding branching fraction is predicted 
to be $\sim4 \times 10^{-5}$.
The $B_s^0 \to \gamma \gamma$ decay is expected to proceed via a
penguin annihilation diagram and to have a branching
fraction in the range $(0.5-1.0)\times 10^{-6}$. 
However, the $B_s^0 \to \gamma \gamma$ decay branching fraction
is sensitive to some Beyond the Standard Model contributions and can be 
enhanced by about an order of magnitude;
such enhanced values are not far from the sensitivity expected 
in this analysis.

The three-dimensional (two-dimensional) unbinned extended 
maximum likelihood fit to $M_{\rm bc}$, $\Delta E$
and cos$\theta_\phi^h$ ($M_{\rm bc}$ and $\Delta E$)
is performed for $B_s^0 \to \phi \gamma$ ($B_s^0 \to \gamma \gamma$)
decay to extract the signal yield.
Fig.~2 shows the $M_{\rm bc}$ and $\Delta E$ projections
of the data.
The points with error bars represent data, the thick solid curves are 
the fit functions, the thin solid curves are the signal functions, 
and the dashed curves show the continuum contribution.
On the $M_{\rm bc}$ figure, signals from $B_s^0 \bar{B}_s^0$,
$B_s^* \bar{B}_s^0$, and $B_s^* \bar{B}_s^*$ appear from left to right.
On the $\Delta E$ figure, due to the requirement $M_{\rm bc} > 5.4\,$GeV/$c^2$
only the $B_s^* \bar{B}_s^*$ signal contributes.

\begin{figure}
\begin{center}
\psfig{figure=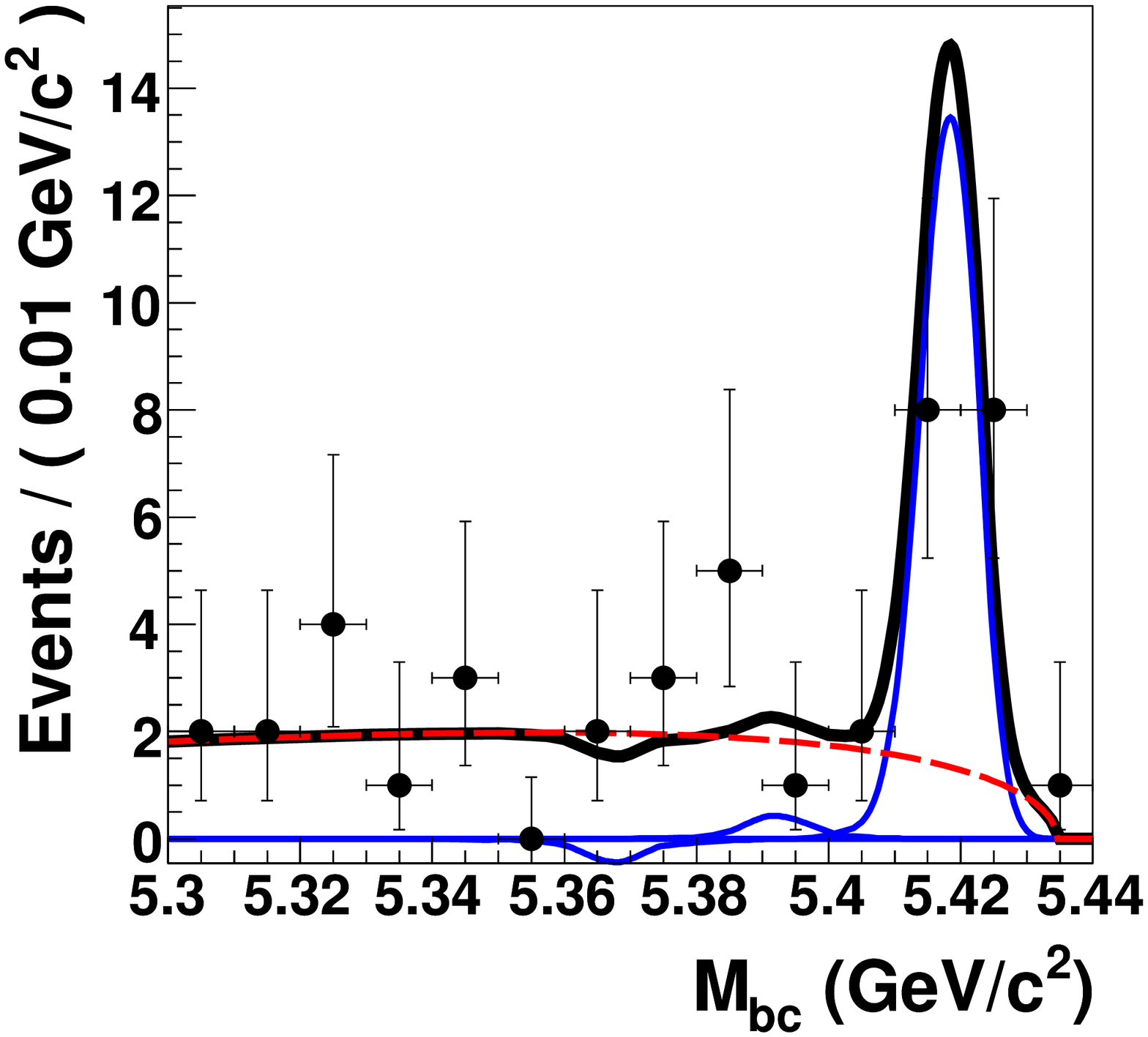,height=1.45in}\psfig{figure=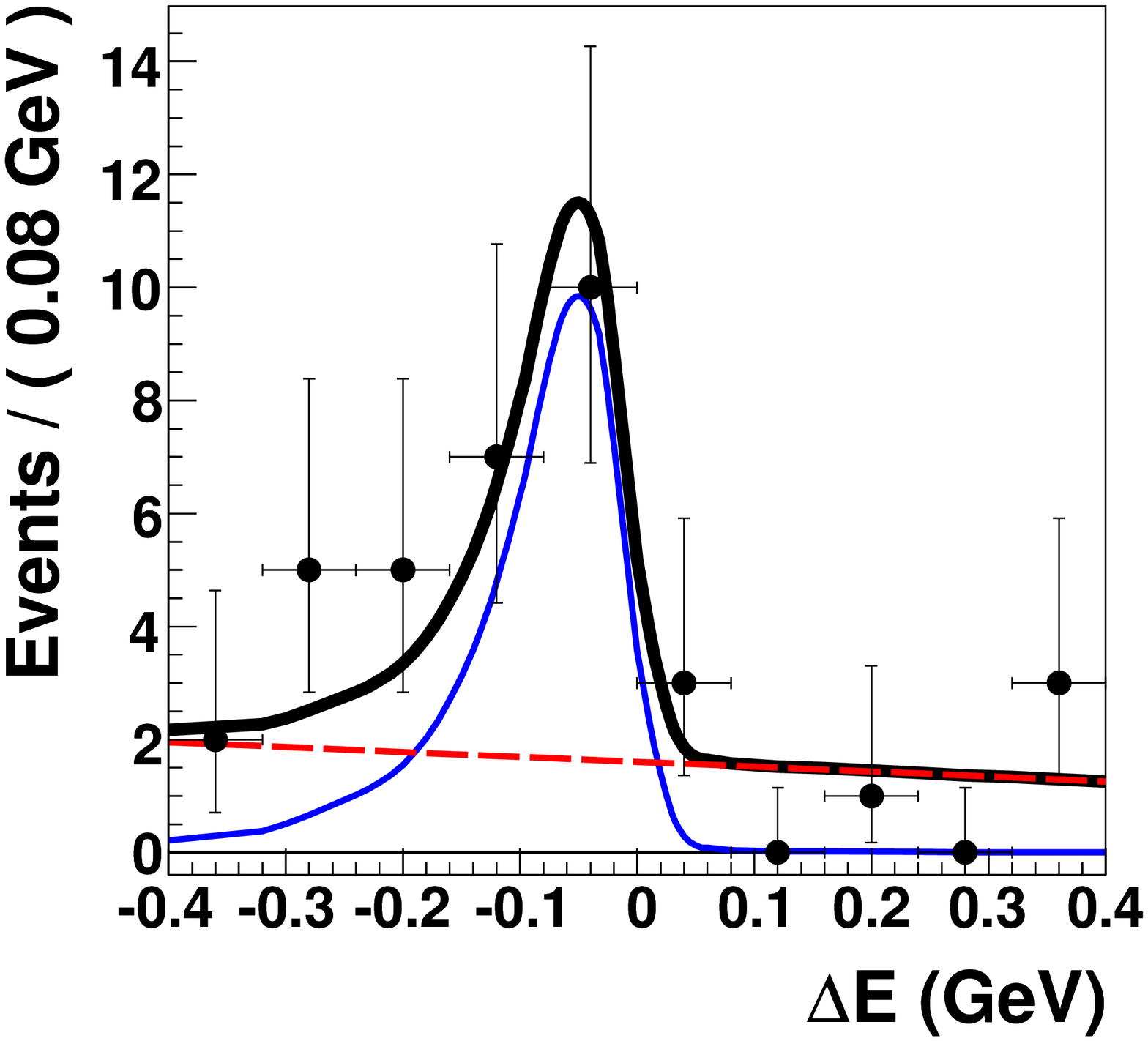,height=1.45in}\psfig{figure=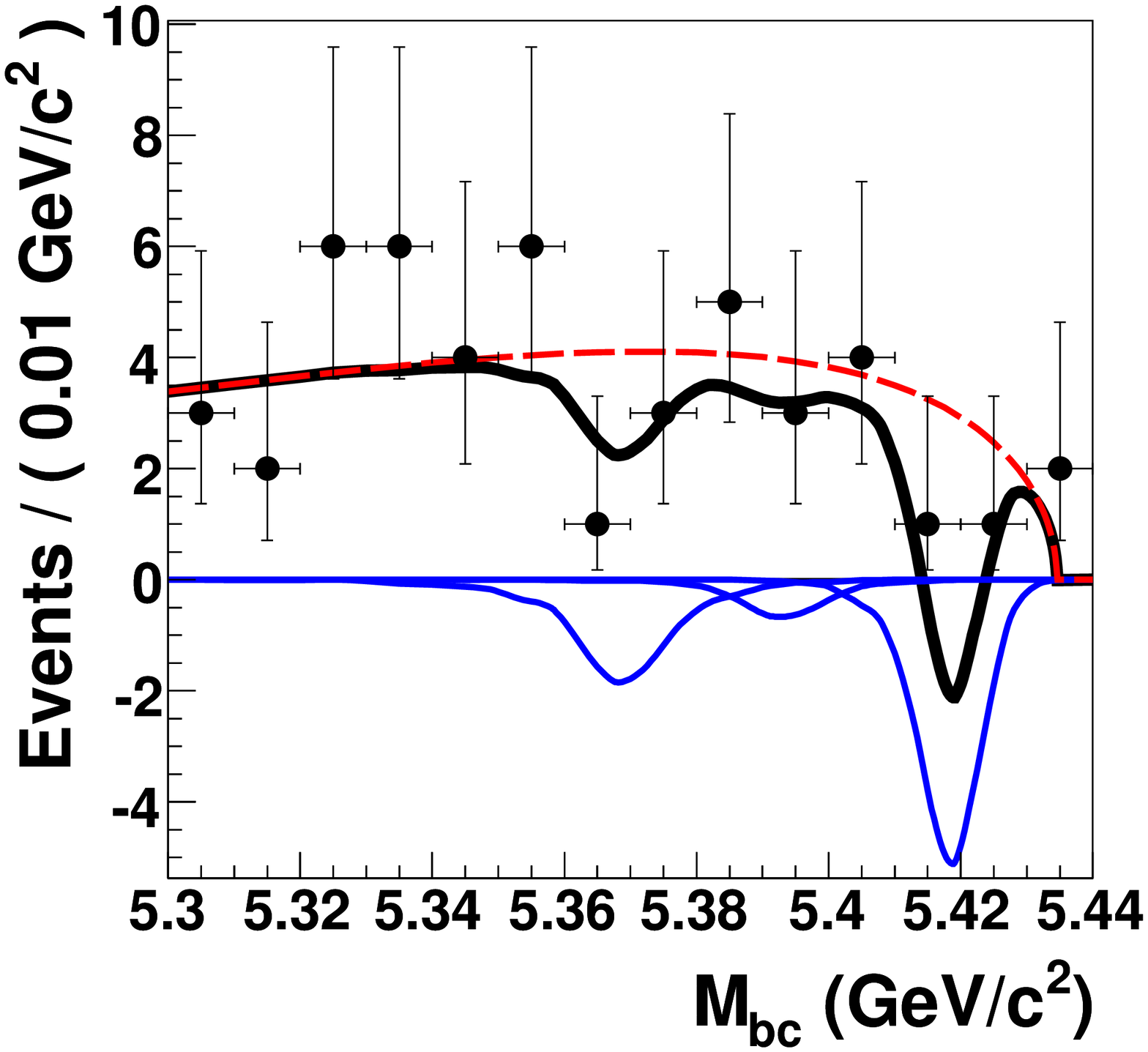,height=1.45in}\psfig{figure=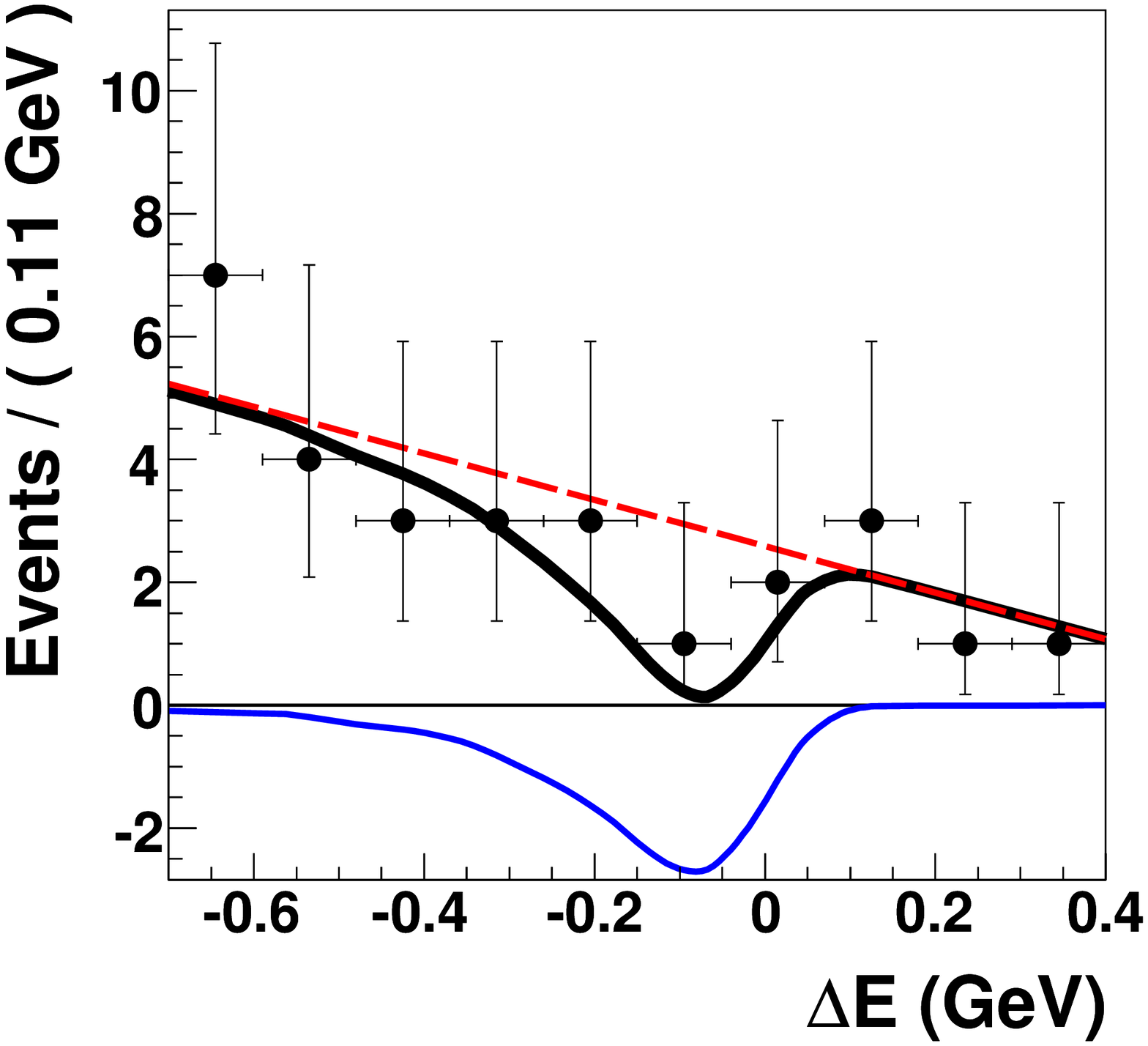,height=1.45in}
\end{center}
\vspace{-0.3cm}
\caption{\vspace{-0.1cm}The $M_{\rm bc}$ and $\Delta E$ projections for the $B_s^0 \to \phi \gamma$ (left) and $B_s^0 \to \gamma \gamma$ (right) modes.
\label{fig:dspia}}
\end{figure}

A clear signal is seen in the $B_s^0 \to \phi \gamma$ mode.
This radiative decay is observed for the
first time and the branching fraction 
${\cal B} (B_s^0 \to \phi \gamma) = (5.7^{+1.8}_{-1.5} {\rm (stat.)} ^{+1.2}_{-1.1} {\rm (syst.)}) \times 10^{-5}$ 
is measured. The obtained value is in agreement
with the SM predictions. No significant signal is observed in the
$B_s^0 \to \gamma \gamma$ mode, and an upper limit at the $90\%$ C.L.
of ${\cal B} (B_s^0 \to \phi \gamma) < 8.7 \times 10^{-6}$ is set.

\subsection{Observation of $B_s^0 \to J/\psi\,\phi$ and $B_s^0 \to J/\psi\,\eta$ decays}\label{subsec:jpsp}

Preliminary results on the 
$B_s^0 \to J/\psi\,\phi$ and $B_s^0 \to J/\psi\,\eta$ decay branching fraction
measurements are obtained by Belle.
The electron mode $J/\psi \to e^+ e^-$ and the muon mode 
$J/\psi \to \mu^+ \mu^-$ are used to reconstruct $J/\psi$ mesons.
The $B_s^0 \to J/\psi\,\phi$ decay branching fraction is measured to be 
$(1.12 \pm 0.25^{+0.21}_{-0.24}) \times 10^{-3}$ and
$(1.18 \pm 0.25^{+0.22}_{-0.25}) \times 10^{-3}$ using the electron
and muon modes, respectively.

The $B_s^0 \to J/\psi\,\eta$ decay branching fraction is measured
using the $\eta \to \gamma \gamma$ and $\eta \to \pi^+ \pi^- \pi^0$
modes to reconstruct $\eta$ mesons. The combined value 
${\cal B} (B_s^0 \to J/\psi\,\eta) = (3.69 \pm 0.95^{+0.65}_{-0.95}) \times 10^{-4}$ is obtained, which is about 3 times smaller than that for the 
$B_s^0 \to J/\psi\,\phi$ decay.
This ratio agrees with a rough estimate obtained within
the quark model, where the $s\bar{s}$ part of the $\eta$ meson wave 
function is one third, in contrast to the fully $s\bar{s}$ content of 
$\phi$ mesons.

\subsection{First measurement of $B_s^0 \to X^+ \ell^- \nu$ decay}\label{subsec:semi}

The correlated production of a $D_s^+$ meson and a same-sign
lepton at the $\Upsilon$(5S) resonance is used in this analysis 
to measure ${\cal B}(B_s^0 \rightarrow X^+ \ell^- \nu)$.
$D_s^+$ candidates are reconstructed in a clean $\phi \pi^+$ mode.
Neither the $c\bar{c}$ continuum nor $B^{(*)}\bar{B}^{(*)}$ states
(except for a small contribution due to $\sim 19\%$ $B^0$ mixing effect) 
can result in a same-sign $c$-quark (i.e.,\ $D_s^+$ meson) and 
primary lepton final state. 

Finally, we obtained the semileptonic branching fractions:
\vspace{-0.15cm}
\begin{eqnarray}
\nonumber
{\cal B}(B_s^0 \rightarrow X^+ e^- \nu)\, = (10.9 \pm 1.0 \pm 0.9)\% \\
{\cal B}(B_s^0 \rightarrow X^+ \mu^- \nu)\, = (9.2 \pm 1.0 \pm 0.8) \% \\
\nonumber
{\cal B}(B_s^0 \rightarrow X^+ \ell^- \nu)\, = (10.2 \pm 0.8 \pm 0.9)\% ,
\vspace{-0.25cm}
\end{eqnarray}
\noindent
where the latter one represents an average over electrons and muons.
The results are preliminary.
The obtained branching fractions can be compared with the PDG value
${\cal B}(B^0 \rightarrow X^+ \ell^- \nu)\, = (10.33 \pm 0.28)\%$ \cite{pdg},
which is theoretically expected to be approximately the same,
neglecting a small possible lifetime difference and
small corrections due to electromagnetic and
light quark mass difference effects.

\section{Study of \boldmath{$\Upsilon$(5S)} decays}

\subsection{Observation of $\Upsilon{\rm (5S)} \to \Upsilon {\rm (1S)}\, \pi^+\pi^-$ and $\Upsilon{\rm (5S)} \to \Upsilon{\rm (2S)}\, \pi^+\pi^-$ decays}\label{subsec:uone}

The production of $\Upsilon{\rm (1S)}\, \pi^+\pi^-$, 
$\Upsilon{\rm (2S)}\, \pi^+\pi^-$, $\Upsilon{\rm (3S)}\, \pi^+\pi^-$,
and $\Upsilon{\rm (1S)}\, K^+K^-$ final states in a 21.7\,fb$^{-1}$
data sample obtained at $e^+ e^-$ collisions with CM energy near the peak 
of the $\Upsilon$(5S) resonance \cite{bups} has been studied by Belle.
Final states with two opposite-sign muons and two opposite-sign
pions (or kaons) are selected. 
Signal candidates are identified using the kinematic variable
$\Delta M$, defined as the difference between $M(\mu^+\mu^-\pi^+\pi^-)$
or $M(\mu^+\mu^-K^+K^-)$ and $M(\mu^+\mu^-)$ for pion or kaon modes.

The obtained branching fractions and partial widths are
given in Table 1. For comparison, the partial widths
for similar transitions from $\Upsilon$(2S), $\Upsilon$(3S), or $\Upsilon$(4S)
are also shown. The $\Upsilon$(5S) partial widths
(assuming that the signal events are solely due to the $\Upsilon$(5S)
resonance) are found to be in the range (0.52-0.85)\,MeV, that is
more than 2 orders of magnitude larger than the corresponding
partial widths for $\Upsilon$(2S), $\Upsilon$(3S), or $\Upsilon$(4S) decays. 
The unexpectedly large $\Upsilon$(5S) partial widths disagree with 
the expectation for a pure $b\bar{b}$ state, unless there is a new mechanism
to enhance the decay rates.

\renewcommand{\arraystretch}{1.2}
\begin{table}[t]
\caption{The branching fractions (${\cal B}$) and the partial widths ($\Gamma$) for $\Upsilon{\rm (nS)} \to \Upsilon{\rm (mS)}\, h^+h^-$ processes.}
\vspace{0.2cm}
\begin{center}
\begin{tabular}{|l|c|c|l|c|}
\hline
$\Upsilon{\rm (5S)} \to$ & ${\cal B}$ ($\%$) & $\Gamma$ (MeV) &  & $\Gamma$ (MeV) \\
\hline
$\Upsilon{\rm (1S)}\, \pi^+\pi^-$ & $0.53 \pm 0.03 \pm 0.05$ & $0.59 \pm 0.04 \pm 0.09$ & $\Upsilon{\rm (2S)} \to \Upsilon{\rm (1S)}\, \pi^+\pi^-$ & 0.006 \\
$\Upsilon{\rm (2S)}\, \pi^+\pi^-$ & $0.78 \pm 0.06 \pm 0.11$ & $0.85 \pm 0.07 \pm 0.16$ & $\Upsilon{\rm (3S)} \to \Upsilon{\rm (1S)}\, \pi^+\pi^-$ & 0.0009 \\
$\Upsilon{\rm (3S)}\, \pi^+\pi^-$ & $0.48^{+0.18}_{-0.15} \pm 0.07$ & $0.52^{+0.20}_{-0.17} \pm 0.10$ & $\Upsilon{\rm (4S)} \to \Upsilon{\rm (1S)}\, \pi^+\pi^-$ & 0.0019 \\
$\Upsilon{\rm (1S)}\, K^+K^-$ & $0.061^{+0.016}_{-0.014} \pm 0.010$ & $0.067^{+0.017}_{-0.015} \pm 0.013$ & & \\
\hline
\end{tabular}
\end{center}
\end{table}

\subsection{Study of $\Upsilon$(5S) decays to $B^0$ and $B^+$ mesons}

The $\Upsilon$(5S) decays to channels with $B^+$ and $B^0$ mesons 
are studied using the 23.6\,fb$^{-1}$ data sample 
obtained at the $\Upsilon$(5S) with the Belle detector.
The reported results are preliminary.
As discussed above, at the $\Upsilon$(5S) energy a $b\bar{b}$ quark pair 
can be hadronized in various final states, such as
two-body $B_s^0$ channels, two-body $B^{+/0}$ channels,
three-body channels $B\bar{B}\,\pi$, $B\bar{B}^\ast\,\pi$, 
$B^\ast\bar{B}\,\pi$, $B^\ast\bar{B}^\ast\,\pi$, and
four-body channel $B\bar{B}\,\pi \pi$.
Generally a $b\bar{b}$ quark pair can also hadronize 
to a bottomonium state accompanied by $\pi$, $K$ or $\eta$ mesons.
 
The $B^+ \to J/\psi K^+$, $B^0 \to J/\psi K^{*0}$,
$B^+ \to \bar{D}^0 \pi^+$ and $B^0 \to D^- \pi^+$ decays are fully
reconstructed and treated simultaneously
to measure the $B^+$ and $B^0$ production rates per $b\bar{b}$ event.
The rates are $f(B^+) = (67.7 \pm 3.6 \pm 4.8)\%$ and 
$f(B^0) = (70.4^{+5.2}_{-5.1} \pm 6.2)\%$,
with the average value of $f(B^{+/0}) = (68.6^{+3.0}_{-2.9} \pm 5.0)\%$.

Assuming equal rates to $B^+$ and $B^0$ mesons
in all channels produced at the $\Upsilon$(5S) energy,
we measure the fractions for $b\bar{b}$ event transitions to
the two-body and multi-body channels with $B^{+/0}$ meson pairs,
$f(B\bar{B}) = (5.1 \pm 0.9 \pm 0.4)\,\%$,
$f(B\bar{B}^*+B^*\bar{B}) = (12.6 \pm 1.2 \pm 1.0)\,\%$,
$f(B^*\bar{B}^*) = (34.7 \pm 1.8 \pm 2.7)\,\%$, and
$f(B^{(*)}\bar{B}^{(*)}\pi(\pi)) = (17.0\,^{+1.6}_{-1.5} \pm 1.2)\,\%$.
The two-body channel fractions are in a reasonable
agreement with theoretical predictions. 
The multi-body channels are observed for the first time and 
the obtained fraction is unexpectedly large.

\end{document}